\documentclass[journal=jacsat,manuscript=article]{achemso}

\usepackage{chemformula} 
\usepackage[utf8x]{inputenc} 
\usepackage[T1]{fontenc} 
\usepackage{mathrsfs}
\usepackage{appendix}
\usepackage{hyperref}
\usepackage{color}
\usepackage[normalem]{ulem}
\usepackage{dcolumn}
\usepackage{adjustbox}
\usepackage{tabularx}
\usepackage{tikz}
\usepackage{newfloat}
\usepackage{physics}
\usepackage{program}
\usetikzlibrary{arrows,shapes,positioning,shadows,trees}

\author{Arif Ullah}
\email{ua2024@xmu.edu.cn}
\author{Pavlo O. Dral}
\email{dral@xmu.edu.cn}
\affiliation[]
{State Key Laboratory of Physical Chemistry of Solid Surfaces, Fujian Provincial Key Laboratory of Theoretical and Computational Chemistry, Department of Chemistry, and College of Chemistry and Chemical Engineering, Xiamen University, Xiamen 361005, China}
\title[]
  {One-shot Trajectory Learning of Open Quantum Systems Dynamics}

\begin{document}

\begin{abstract}
    Nonadiabatic quantum dynamics are important for understanding light-harvesting processes, but their propagation with traditional methods can be rather expensive. Here we present a one-shot trajectory learning approach that allows to directly make ultra-fast prediction of the entire trajectory of the reduced density matrix for a new set of such simulation parameters as temperature and reorganization energy. The whole 10~ps long propagation takes 70 milliseconds as we demonstrate on the comparatively large quantum system, the Fenna--Matthews--Olsen (FMO) complex. Our approach also significantly reduces time and memory requirements for training.
\begin{figure}[htb]
     \centering
     \includegraphics[width=5.08cm]{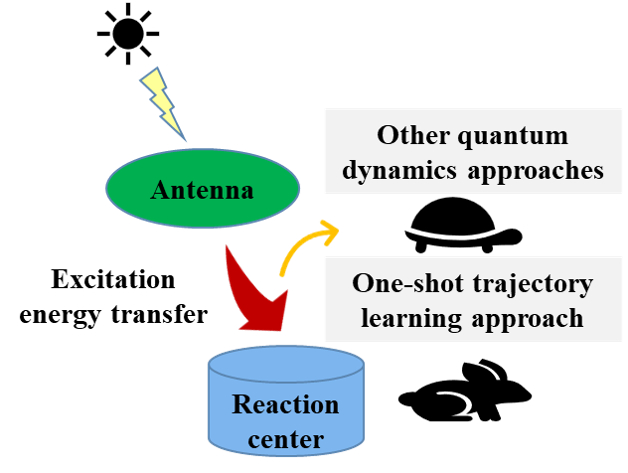}
 \end{figure}
\end{abstract}

\maketitle

Life directly or indirectly depends on the light-harvesting processes. Plants, algae, and photosynthetic bacteria possess highly efficient light-harvesting systems, transforming light energy into the required form of the chemical energy. In natural light-harvesting systems, an antenna absorbs photons, which are transferred to a reaction center through an exciton transfer complex. The research into highly efficient funneling of excitation energy in these natural complexes is also important for the design of highly efficient organic solar devices.\cite{tronrud2009structural} The Fenna--Matthews--Olsen (FMO) complex which is found in green sulfur bacteria\cite{adolphs2006proteins, karafyllidis2017quantum} is one of the most well-studied exciton transfer complexes because of its near-unity efficiency, small size, and simplicity.

Many traditional quantum dynamics methods are explored to study the excitation energy transfer in FMO complex.\cite{ishizaki2009theoretical, collini2010coherently, milder2010revisiting,schmidt2011eighth, olbrich2011theory,chenu2015coherence,he2021commutator,liu2021unified} All of them are recursive, i.e., require iterative dynamics propagation and at the same time, computationally expensive. Recently, ML has been successfully applied to accelerate quantum dynamics.\cite{bandyopadhyay2018applications, banchi2018modelling, yang2020applications,herrera2021convolutional,ullah2021speeding,lin2021simulation,wu2021forecasting,hase2017heom,ullah2021predicting,choi2022learning, kadupitiya2022solving,lin2022auto} ML-based approaches adopted in the literature so far, can be divided into two categories: recursive\cite{herrera2021convolutional,ullah2021speeding,lin2021simulation,wu2021forecasting} and non-recursive\cite{hase2017heom, ullah2021predicting} approaches. ML-based recursive approaches where dynamics at some time $t$ depends on the dynamics of previous time-steps, are quite successful but there are some downsides of them; first, iterative propagation is inherently slow and may lead to error accumulation and second, they need a short-time trajectory generated with traditional quantum dynamics methods to initiate trajectory propagation. 

One of the non-recursive approaches was learning outcome of dynamics (transfer time and efficiency) rather than dynamics.\cite{hase2017heom} Recently, we have proposed an alternative non-recursive, artificial intelligence-based quantum dynamics (AI-QD) approach, learning dynamics trajectories as a function of time and simulation parameters: the reorganization energy $\lambda$, characteristic frequency $\gamma$, and temperature $T$.\cite{ullah2021predicting} This makes all time-steps independent from each other which allows us to predict system's state at any arbitrary time, without the need of propagating the trajectory.

Despite a significant speed-up of dynamics propagation with our AI-QD approach, generating the entire trajectory still needs the evaluation of ML function at each required time-step of the trajectory, making it relatively slow. In addition, the time and memory requirements for training are rather high because of the way the data for supervised learning is prepared: AI-QD is trained on as many points as there are time-steps in all training trajectories, i.e., training set contains millions of points.\cite{ullah2021predicting}

\begin{figure}
    \centering
    \includegraphics[width=\textwidth]{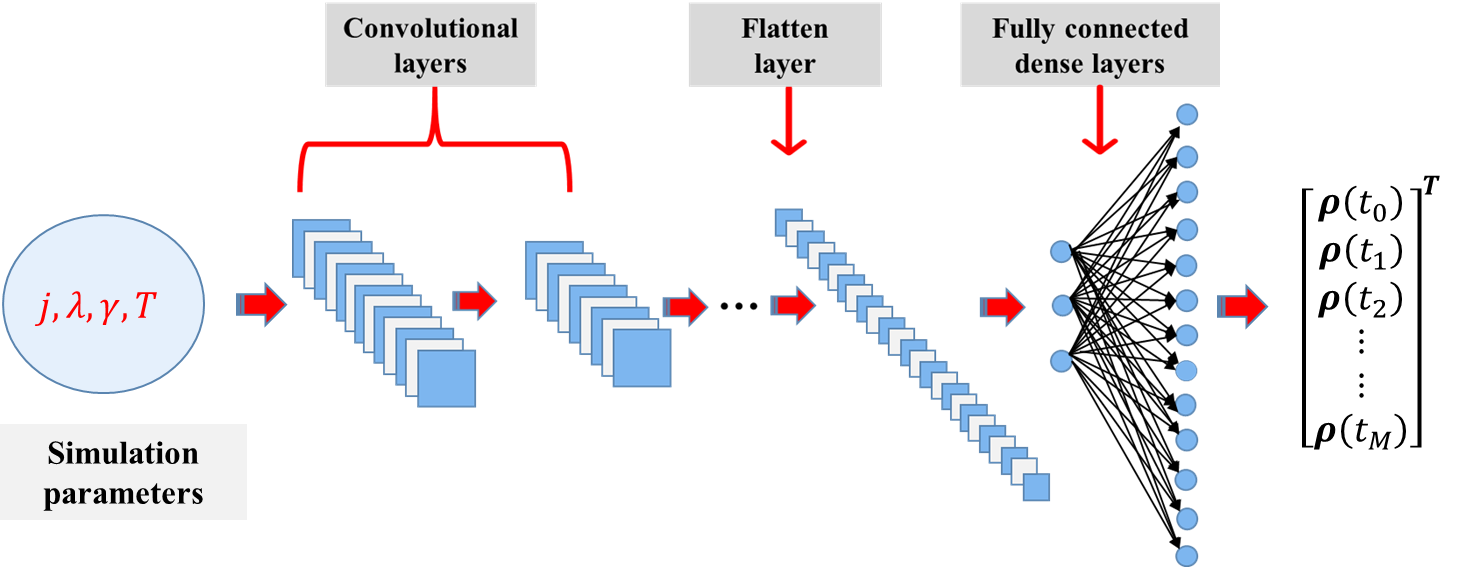}
    \caption{Predicting the entire $t_M = 10$~ps long-time evolution of reduced density matrix $\boldsymbol{\rho}(t)$ with the one-shot trajectory learning approach which takes as input simulation parameters. The simulation parameters investigated in this work are the reorganization energy $\lambda$, characteristic frequency $\gamma$, and temperature $T$. An additional input $j$ represents the label of a site with initial excitation. In this work we use a one-dimensional convolutional neural network.}
    \label{fig:sketch}
\end{figure}

Here we propose a one-shot trajectory learning (OSTL) quantum dynamics approach, which is a much more efficient strategy for both learning and trajectory propagation than AI-QD. Similar to AI-QD, OSTL takes the simulation parameters as input but then in contrast to AI-QD, OSTL directly evaluates the entire trajectory as a multi-output prediction by a one-dimensional (1D) convolutional neural network (CNN) (Fig.~\ref{fig:sketch}). The time range of the entire trajectory is set by the maximum run time of the training trajectories. Our new strategy only requires up to 70 milliseconds for 10~ps long trajectory for the reduced density matrix $\boldsymbol{\rho}(t)$ in the seven sites FMO complex\cite{adolphs2006proteins, karafyllidis2017quantum} which was also used in previous ML studies.\cite{hase2017heom, ullah2021predicting} OSTL also has significantly reduced computational requirements for training as a multi-output 1D CNN is trained on the number of points equal to the number of training trajectories.

For demonstration, we use the same (10~ps long) training trajectories propagated with the traditional quantum dynamics method (Local thermalising Lindblad master equation (LTLME)\cite{Mohseni2008Environ} and Frenkel exciton Hamiltonian of seven sites FMO\cite{ishizaki2009unified}) and the same simulation parameters as used in the AI-QD demonstration (see Methods and Ref.~\citenum{ullah2021predicting}). Each traditional training trajectory of the reduced density matrix $\boldsymbol{\rho}(t)$ of the FMO complex is transformed into a single training point (Fig.~\ref{fig:mtpbnr}). The single training point contains a target vector $[\boldsymbol{Y}(t_0)$,$\boldsymbol{Y}(t_1)$, $\dots$, $\dots$, $\boldsymbol{Y}(t_M)]$  where $\boldsymbol{Y}(t) = [\rho_{11}(t)$, $\dots$, $\rho_{17}(t)$, $\rho_{22}(t)$, $\dots$, $\rho_{27}(t)$, $\rho_{33}(t)$, $\dots$, $\rho_{37}(t)$,$\dots$, $\dots$, $\rho_{66}(t)$, $\rho_{67}(t)$ ,$\rho_{77}(t)]$ with dimension $N = 49 = $ number of sites + number of upper off-diagonal terms $\times$ 2. As $\rho_{nm}(t) = \rho_{mn}^*(t) \, (n \neq m)$, we only consider the upper diagonal terms $\rho_{nm}(t)\, (n \le m)$ and as the off-diagonal terms are complex, we learn the real and imaginary parts independently. For training, in time-window $0\dots 2.5$~ps, we sample $\boldsymbol{\rho}(t)$ at each 5~fs while beyond 2.5~ps, we sample $\boldsymbol{\rho}(t)$ with 25~fs, totalling to 801 time-steps in the single vector-label.
 In OSTL approach, as a traditional trajectory transferred into a single training point, the number of labelled points used for supervised learning is equal to the number of training trajectories. It is many orders of magnitude smaller than in the previously developed recursive and non-recursive single-time-step ML approaches. In the latter, each trajectory was transformed to a large number of labelled points used for training dependent on the number of time-steps. This reduction leads to significantly faster training and lower memory requirements.

\begin{figure}[!t]
    \centering
    \begingroup
    \small
\text{One unsupervised trajectory of reduced density matrix $\boldsymbol{\rho}(t)$} \\
\text{\underline{with initial excitation on site-1 or site-6 represented by label $j$}} \\    
\begin{equation*}
\begin{bmatrix}
\rho^{(k)}_{11}(t_0) & \dots & \rho^{(k)}_{17}(t_0) \\
\vdots  & \vdots & \vdots \\
\rho^{(k)}_{71}(t_0) & \dots & \rho^{(k)}_{77}(t_0)
\end{bmatrix}
\begin{bmatrix}
\rho^{(k)}_{11}(t_1) & \dots & \rho^{(k)}_{17}(t_1) \\
\vdots  & \vdots & \vdots \\
\rho^{(k)}_{71}(t_1) & \dots & \rho^{(k)}_{77}(t_1)
\end{bmatrix}
\dots
\dots
\begin{bmatrix}
\rho^{(k)}_{11}(t_M) & \dots & \rho^{(k)}_{17}(t_M) \\
\vdots  & \vdots & \vdots \\
\rho^{(k)}_{71}(t_M) & \dots & \rho^{(k)}_{77}(t_M)
\end{bmatrix}
\end{equation*}
\begin{equation*}
      \Bigg\Downarrow
\end{equation*}
\text{\underline{A single training point}} \\ 
\begin{equation*} 
\arraycolsep=1.5pt\def\arraystretch{2.2}
\begin{array}{cccc|cccccccc}
j &  \lambda^{(k)} &  \gamma^{(k)} & T^{(k)} & \boldsymbol{Y}(t_0) & \boldsymbol{Y}(t_1)& \dots& \boldsymbol{Y}(t_{i-1}) & \boldsymbol{Y}(t_i)&  \boldsymbol{Y}(t_{i+1})&\dots& \boldsymbol{Y}(t_M) \\
\hline 
&  & \text{Input}   &    &  &  &    & \text{Target}   &  &  &  & 
\end{array}
\end{equation*}
\begin{equation*}
    \boldsymbol{Y}(t) = \rho_{11}(t), \dots, \rho_{17}(t), \rho_{22}(t), \dots, \rho_{27}(t), \rho_{33}(t), \dots, \rho_{37}(t),\dots, \dots, \rho_{66}(t), \rho_{67}(t), \rho_{77}(t)
\end{equation*}
\endgroup
    \caption{Transformation of the $k$th unsupervised trajectory of the reduced density matrix $\boldsymbol{\rho}(t)$ propagated with the traditional quantum dynamics method from initial time $t_0$ up to simulation time $t_M$ into one training point for the OSTL model. We include $j$ as a label to distinguish between the initial excitation on site-1 and site-6. Simulation parameters $\lambda$, $\gamma$ and $T$ are the reorganization energy, characteristic frequency, and temperature, respectively. As $\rho_{nm}(t) = \rho_{mn}^*(t) \, (n \neq m)$, we consider only the upper diagonal terms $\rho_{nm}(t)\, (n \le m)$ and as the off-diagonal terms are complex, we learn the real and imaginary parts independently.}
    \label{fig:mtpbnr}
\end{figure}

\begin{figure}[!t]
    \centering
    \includegraphics[width=\textwidth]{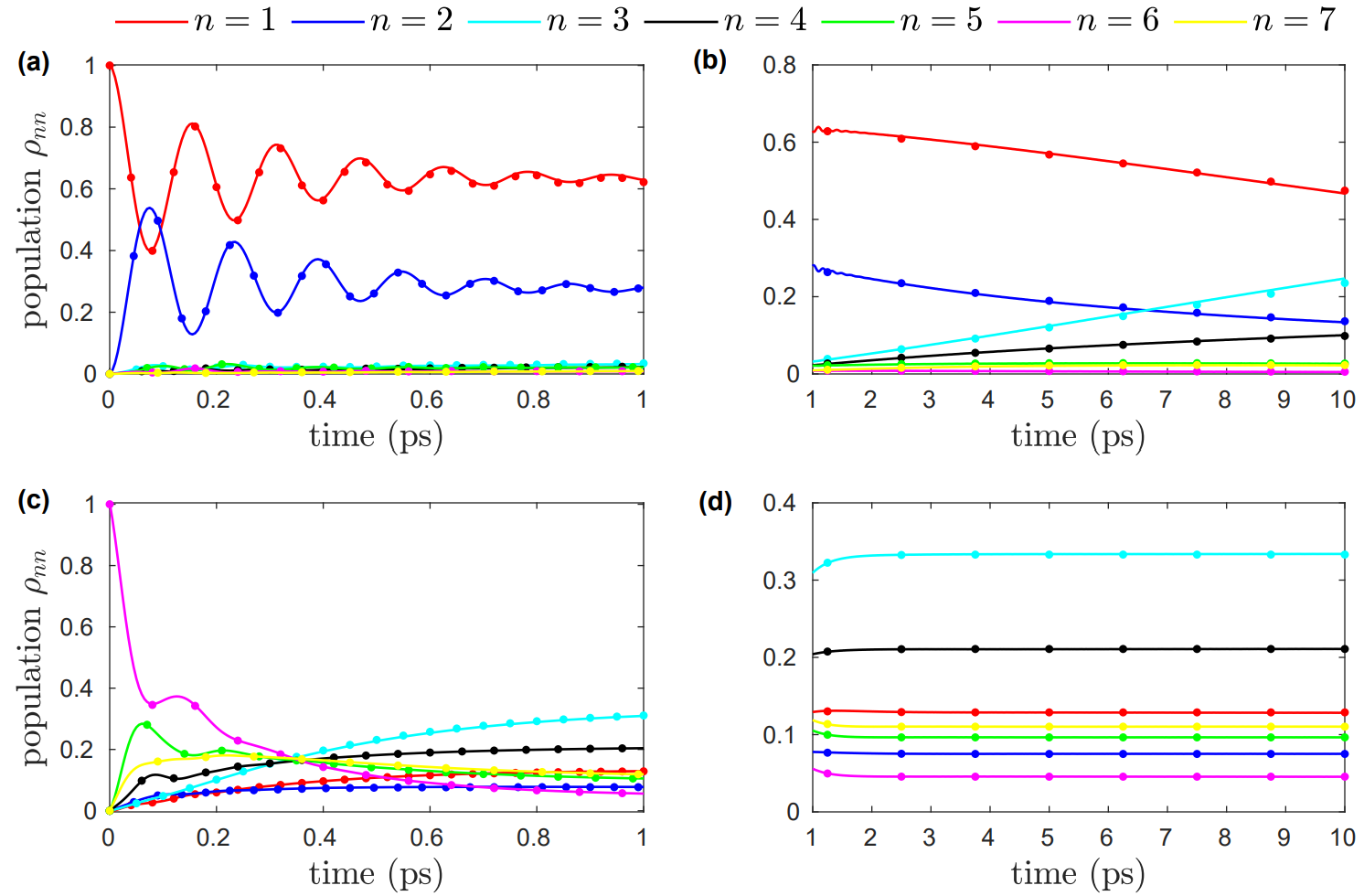}
    \caption{Population of the seven sites in FMO complex as a function of time.  In (a) and (b) the initial excitation is on site-1 and other parameters are $\gamma = 300$, $\lambda =10$, $T=50$. In (c) and (d), the initial excitation is on site-6 and other parameters are $\gamma = 75$, $\lambda =40$, $T=290$. In (a) and (c), we show 1~ps dynamics and dynamics beyond 1~ps is shown in (b) and (d). We compare our results to the reference LTLME method (dots). Dynamics of the corresponding off-diagonal terms are shown Fig.~\ref{fig:coherence}. In our calculations, $\gamma$ and $\lambda$ are considered in the units of cm$^{-1}$, while $T$ is in the units of K.}
    \label{fig:eet}
\end{figure}

\begin{figure}[!t]
    \centering
    \includegraphics[width=\textwidth]{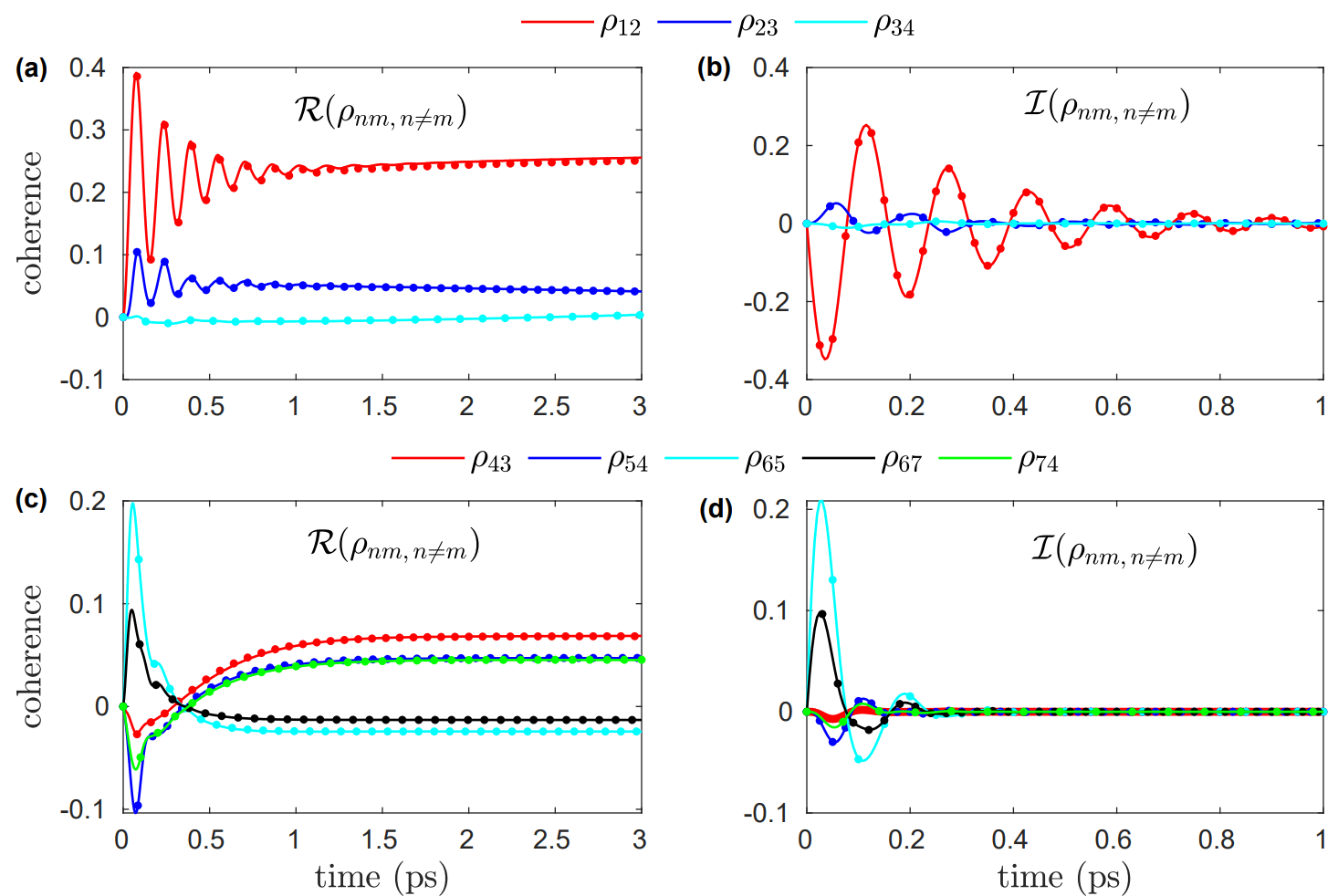}
    \caption{Electronic coherence or the off-diagonal terms of reduced density matrix $\boldsymbol{\rho}(t)$ as a function of time. In (a) and (b), we respectively show the real and imaginary parts of the prominent off-diagonal terms for Fig.~\ref{fig:eet}(a),(b) where $\gamma = 300$, $\lambda =10$, $T=50$ with initial excitation on site-1. (c) and (d) respectively show the real and imaginary parts of the prominent off-diagonal terms for Fig.~\ref{fig:coherence}, where the initial excitation is on site-6 and other parameters are $\gamma = 75$, $\lambda =40$, $T=290$. We compare our results to the reference LTLME method (dots). In our calculations, $\gamma$ and $\lambda$ are considered in the units of cm$^{-1}$, while $T$ is in the units of K.}
    \label{fig:coherence}
\end{figure}

We evaluate the performance of the OSTL approach for excitation energy transfer in FMO complex by training on 1000 trajectories chosen based on farthest point sampling\cite{dral2019mlatom} (500 trajectories for each case of initial excitation as described previously\cite{ullah2021predicting}). The chosen 1000 training trajectories make up only ca. 25\% of the whole data set (see "Methods" and Ref.~\citenum{ullah2021predicting}) and we evaluate its performance on ca. 70\% remaining trajectories of the data set (2760 test trajectories; 1380 trajectories for each site with initial excitation). We trained multi-output 1D CNN for 10 thousand epochs and saved the model with the smallest validation loss (mean squared error of $2.57\cdot 10^{-7}$) evaluated for 200 trajectories which make up ca. 5\% of the whole data set (not included in the training set and chosen by the farthest point sampling). This best ML model was used to propagate 10~ps long-time dynamics for the test trajectories predicting the exciton populations and coherence for seven sites (Figs.~\ref{fig:eet} and \ref{fig:coherence}). The OSTL trajectories for the reduced density matrix $\boldsymbol{\rho}(t)$ agree very well with the reference traditional trajectories for both initial excitations on site-1 and site-6 demonstrating that OSTL approach can successfully learn the challenging quantum beating (modulation of amplitudes) in the excitation energy transfer, along with the off-diagonal terms representing the coherence (Figs.~\ref{fig:eet} and \ref{fig:coherence} show trajectories for one random set of simulation parameters, other trajectories are provided as described in the Data availability section).

The mean absolute error (MAE) and root mean square error (RMSE) for both diagonal and off-diagonal terms averaged over 2760 hold-out test trajectories are given in Table~\ref{tab:error}, which are rather small. Prediction of the 10~ps long trajectory takes less than 70 millisecond; for comparison, the same propagation with our previous AI-QD approach\cite{ullah2021speeding} takes ca. 2~min on the same machine (single core of Intel(R) Core(TM) i7-10700 CPUs @ 2.90 GHz). Such a tremendous speed-up in propagating quantum dynamics owes to the parallel computation of all time-steps. In contrast to single-time-step ML-based approaches\cite{herrera2021convolutional,ullah2021speeding,lin2021simulation,wu2021forecasting,ullah2021predicting} where the entire ML architecture is invoked for each time-step, in our proposed OSTL multi-time-steps output approach, only the output layer is different for each time-step. In addition, because of the significant reduction in the number of training points, the training process with 8730 epochs takes only 2~h on 16 cores of Intel(R) Xeon(R) Gold 6226R @ 2.90GHZ CPUs. 
\begin{table}[!htb]
	\caption{Mean absolute error (MAE) and root mean square error (RMSE) averaged over the hold-out test set of 2760 trajectories propagated up to 10~ps. $\mathcal{R}\{\rho_{mn, \, n \neq m}\}$ and $\mathcal{I}\{\rho_{mn, \, n \neq m}\}$ represent the real and imaginary part of the off-diagonal terms, respectively.}
	\centering
	\begin{tabular}{l*{5}c} 
		\hline\hline
		& \multicolumn{1}{c}{diagonal terms} &  & \multicolumn{2}{c}{off-diagonal terms} \\ \cline{2-2}  \cline{4-5}
		Error & \multicolumn{1}{c}{$\rho_{nn}$} & &\multicolumn{1}{c}{$\mathcal{R}\{\rho_{mn, \, n \neq m}\}$} & \multicolumn{1}{c}{$\mathcal{I}\{\rho_{mn, \, n \neq m}\}$}  \\ \hline
		MAE  & 3.9$\cdot10^{-4}$ & & 1.3$\cdot10^{-4}$ & 3.1$\cdot10^{-5}$  \\
		
		RMSE & 5.1$\cdot10^{-4}$ & & 1.8$\cdot10^{-4}$ & 6.1$\cdot10^{-5}$ \\     
		\hline\hline
	\end{tabular}
	\label{tab:error}
\end{table}

In summary, we have proposed a one-shot trajectory learning (OSTL) approach which allows us to extremely fast propagate quantum dynamics. The multi-time-steps output feature of our approach avoids invoking the entire neural network architecture for each time-step, thus significantly speeding-up quantum dynamics propagation. The proposed approach is non-recursive, hence also avoiding the problem of error accumulation and eliminating the need of any short-time trajectory as an input similarly to AI-QD. In contrast to AI-QD, OSTL significantly decreases the number of points (labelled data) used for supervised learning, thus making training much more efficient, which is especially useful in the case of large systems where data can be very large. 
As any supervised ML approach, OSTL requires a generation of the reference data to be trained on, but once trained, it can be exploited for fast propagation in massive quantum dynamics simulations, i.e., by interpolating in the simulation parameter space as shown in Fig.~S1 of the Supporting Information and finding the desired parameters. Because our OSTL approach is neither recursive nor includes time in the input, it cannot predict dynamics beyond the maximum run time of the training trajectories. Such predictions are not the design target of OSTL, while longer-time propagation from shorter trajectories is the design target of the recursive ML approaches\cite{herrera2021convolutional,ullah2021speeding,lin2021simulation,wu2021forecasting}. Hence, both OSTL and recursive ML should be considered complementary approaches and they can be combined when both massive simulations and longer-time propagation are required. Finally, although this study is dedicated to presenting a new methodology, future studies may focus on using more accurate reference approaches than an approximate method LTLME to generate data for training.

\section*{Methods} \label{sec:theory}
In this work, we use multi-output one-dimensional (1D) convolutional neural network (CNN). For training and testing we use the data set of trajectories (10~ps long) propagated for the seven sites FMO complex from the Ref.~\citenum{ullah2021predicting}. In this entire data set including training, validation and test sets, the range of simulation parameters is $\lambda=\{10$, $40$, $70$, $100$, $130$, $160$, $190$, $220$, $250$, $280$, $310\}$~cm$^{-1}$, $\gamma=\{25$, $50$, $75$, $100$, $125$, $150$, $175$, $200$, $225$, $250$, $275$, $300\}$~cm$^{-1}$, $T=\{30$,  $50$, $70$, $90$, $110$, $130$, $150$, $170$, $190$, $210$, $230$, $250$, $270$, $290$, $310\}$~K. Before training, we normalize $\lambda$, $\gamma$ and $T$ using their maximum values as normalization factors, i.e.,  $\lambda = \{\lambda_1, \lambda_2, \lambda_3, \dots \lambda_j\}/ \lambda_{\rm max}$, $\gamma= \{\gamma_1, \gamma_2, \gamma_3, \dots \gamma_k\}/\gamma_{\rm max}$, $T=\{T_1, T_2, T_3,  \dots  T_l\}/T_{\rm max}$. Labels $j$ for sites with initial excitation are 0 and 1 which respectively represent initial excitation on site-1 and site-6. We optimize 1D CNN architecture with the hyperopt library.\cite{Bergstra2015HyperoptAP} Then, we train the 1D CNN using the Keras software package\cite{Keras2015} with the TensorFlow in the backend.\cite{Abadi2016TensorFlowLM} The 1D CNN architecture that we used is given in Table~\ref{tab:ostl} with 0.001 as learning rate and 16 as batch size.

\begin{table*}
\caption{Summary of the optimized neural network architecture with layers, output shape (OS), number of parameters (NP), activation function (AF), number of filters (NF), kernel size (KS) and number of neurons (NN) for the MT-PBNR approach.}
 \begin{tabular}{l*{7}c} 
\hline\hline
  \multicolumn{7}{c}{1D CNN architecture for OSTL approach} \\  \cline{1-7}
 Layers (type) & \multicolumn{1}{c}{OS} & \multicolumn{1}{c}{NP} & \multicolumn{1}{c}{AF}  & \multicolumn{1}{c}{NF} &   \multicolumn{1}{c}{KS} & \multicolumn{1}{c}{NN} &  \\ \hline
1st Conv1D &  (None, 2, 80)  & 320   & relu & 80 & 3 &  - \\

2nd Conv1D  &         (None, 2, 110)      &    26,510  & relu & 110 & 3 & - \\     

3rd Conv1D  &         (None, 2, 80)      &    26,480  & relu & 80 & 3 & - \\

Max-Pooling layer   & (None, 1, 80)    &       0  & -  & - & - & -\\  

Flatten layer        &     (None, 80)       &    0     & -  & - & - & -\\  

1st dense layer        &         (None, 32)       &        2592  & relu &  -& - & 32\\     

2nd dense layer     &         (None, 128)       &        4224 & relu &  - & - & 128\\    

Dense output layer     &         (None, 39,249)         &         5,063,121    & linear  & - & -& 39,249 \\  
\hline\hline
\end{tabular}
\vspace{5pt}\\
\centering Total number of parameters: 5,123,247 \\
\label{tab:ostl}
\end{table*}

\section*{Authors contributions}
A.U. conceived the idea of one-shot trajectory learning for quantum dynamics propagation and developed the method. Both authors designed the project. A.U. did all the implementations, calculations, analysis of data, and wrote the original version of the manuscript. Both authors revised the manuscript.

\section*{Code and Data availability}
The corresponding code and data for peer-review are provided at \\ \href{https://figshare.com/s/42328679bb7391f368cc}{https://figshare.com/s/42328679bb7391f368cc}  
and will be released publicly on the acceptance of this manuscript.

\section*{Acknowledgements}
P.O.D. acknowledges funding by the National Natural Science Foundation of China (No. 22003051), the Fundamental Research Funds for the Central Universities (No. 20720210092), and via the Lab project of the State Key Laboratory of Physical Chemistry of Solid Surfaces.

\begin{suppinfo}
Additional results demonstrating the interpolation and extrapolation in the parameter space and a table showing the mean square errors (MSEs) and root mean square errors (RMSEs) for them.   
\end{suppinfo}

\section*{Competing interests}
The authors declare no competing interests.

\bibliography{ref.bib}

\end{document}


\begin{suppfigure}
     \centering
     \includegraphics[width=\textwidth]{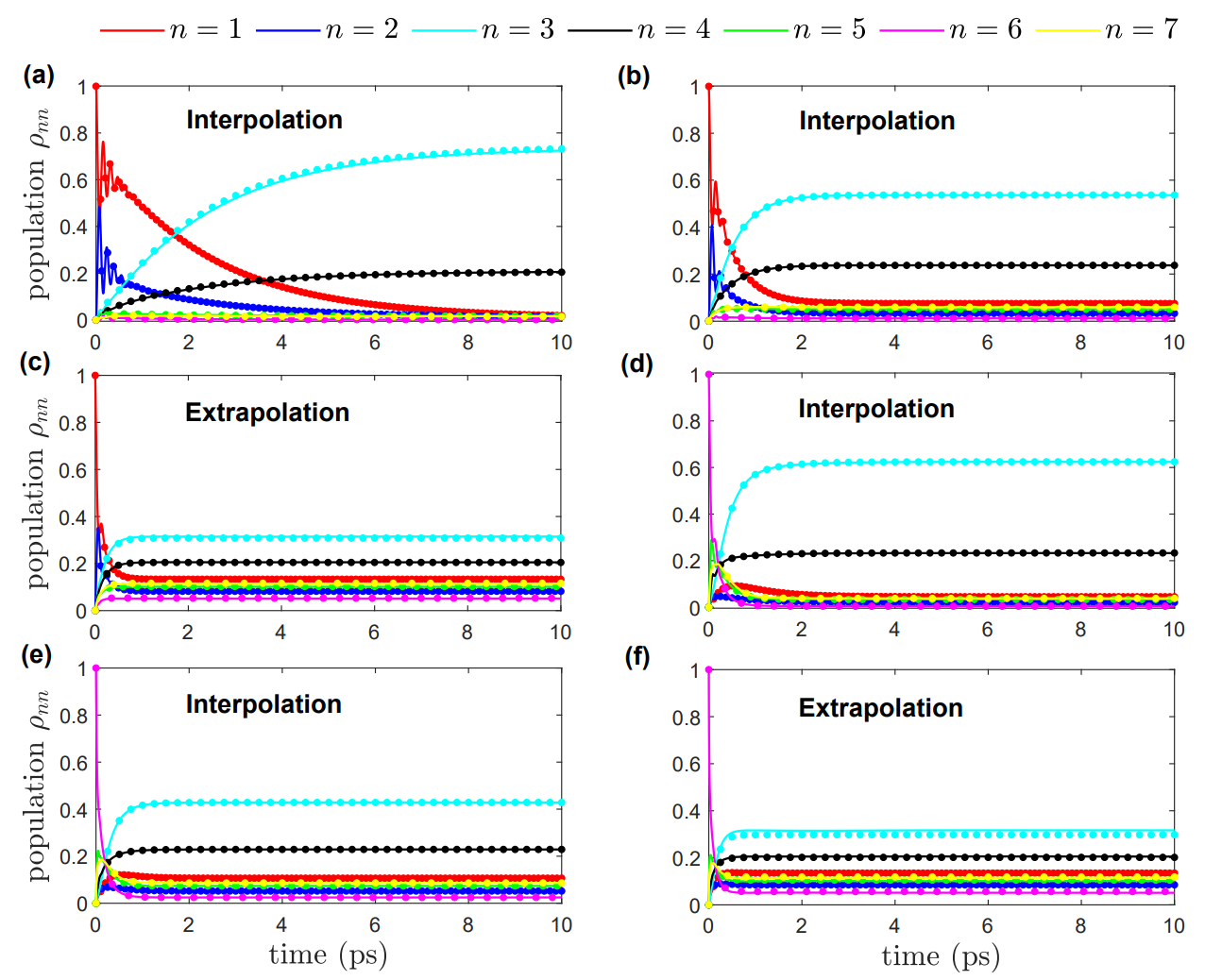}
     \caption{Population dynamics of the seven sites in the FMO complex as a function of time. Results are shown for parameters which do not appear in the training set at all, demonstrating interpolation and extrapolation. With initial excitation on site-1, simulation parameters for (a), (b) and (c) are (a) $\gamma = 65$, $\lambda =35$, $T=55$, (b) $\gamma = 155$, $\lambda =145$, $T=135$ and (c) $\gamma = 330$, $\lambda =330$, $T=330$. In (d), (e) and (f), the initial excitation is on site-6 and other parameters are (d) $\gamma = 80$, $\lambda =85$, $T=100$, (e) $\gamma = 205$, $\lambda =185$, $T=195$ and (f) $\gamma = 350$, $\lambda =350$, $T=350$. The results of AI-QD are compared to the LTLME results (dots). $\gamma$ and $\lambda$ are in the units of cm$^{-1}$, while $T$ is in the units of K. The respective mean square errors (MAEs) and root mean square errors (RMSEs) are given in Table~\ref{tab:inter_extrapolation}.}
     \label{fig:intra_extrpolation}
 \end{suppfigure}
 \begin{supptable*}
\caption{Mean absolute error (MAE) and root mean square error (RMSE) for the test trajectories presented in Fig.~\ref{fig:intra_extrpolation}. $\mathcal{R}\{\rho_{mn, \, n \neq m}\}$ and $\mathcal{I}\{\rho_{mn, \, n \neq m}\}$ represent the real and imaginary parts of the off-diagonal terms, respectively. The IES denotes the initial excited site.}
\begin{adjustbox}{width=\textwidth}
 \begin{tabular}{l*{12}c} 
 \hline\hline
  & \multicolumn{4}{c}{parameters}  &  &\multicolumn{3}{c}{MAE} &  & \multicolumn{3}{c}{RMSE} \\ \cline{2-5} \cline{7-9} \cline{11-13}
 & \multicolumn{1}{c}{IES} & \multicolumn{1}{c}{$\gamma$} & \multicolumn{1}{c}{$\lambda$} & \multicolumn{1}{c}{$T$}   & & \multicolumn{1}{c}{$\rho_{nn}$} &  \multicolumn{1}{c}{$\mathcal{R}\{\rho_{mn, \, n \neq m}\}$} & \multicolumn{1}{c}{$\mathcal{I}\{\rho_{mn, \, n \neq m}\}$} & & \multicolumn{1}{c}{$\rho_{nn}$} &  \multicolumn{1}{c}{$\mathcal{R}\{\rho_{mn, \, n \neq m}\}$} & \multicolumn{1}{c}{$\mathcal{I}\{\rho_{mn, \, n \neq m}\}$} \\ \hline
  \multicolumn{13}{c}{Interpolation in the parameter space} \\
  \hline
Fig.~S1(a) & 1 & 65 & 35 & 55 & & 2.4$\cdot10^{-3}$ &  7.0$\cdot10^{-4}$ & 5.8$\cdot10^{-5}$ & & 2.8$\cdot10^{-3}$ & 8.2$\cdot10^{-4}$ &  1.2$\cdot10^{-4}$\\
Fig.~S1(b) & 1 & 155 & 145 & 135 & & 1.9$\cdot10^{-4}$ &  1.0$\cdot10^{-4}$ & 2.3$\cdot10^{-5}$ & & 2.6$\cdot10^{-4}$ & 1.2$\cdot10^{-4}$ &  4.8$\cdot10^{-4}$\\
Fig.~S1(d) & 6 & 80 & 85 & 100 & & 4.9$\cdot10^{-4}$ &  1.7$\cdot10^{-4}$ & 3.3$\cdot10^{-5}$ & & 6.3$\cdot10^{-4}$ & 2.3$\cdot10^{-4}$ &  7.4$\cdot10^{-5}$\\
Fig.~S1(e) & 6 & 205 & 185 & 195 & & 1.2$\cdot10^{-4}$ &  5.2$\cdot10^{-5}$ & 1.8$\cdot10^{-5}$ & & 1.7$\cdot10^{-4}$ & 7.4$\cdot10^{-5}$ &  3.4$\cdot10^{-5}$\\
\hline\hline
\multicolumn{13}{c}{Extrapolation in the parameter space} \\ \cline{1-13} \\
Fig.~S1(c) & 1 & 330 & 330 & 330 & & 1.9$\cdot10^{-3}$ &  4.6$\cdot10^{-4}$ & 5.2$\cdot10^{-5}$ & & 2.0$\cdot10^{-3}$ & 5.4$\cdot10^{-4}$ &  1.0$\cdot10^{-4}$\\
Fig.~S1(f) & 6 & 350 & 350 & 350 & & 5.7$\cdot10^{-3}$ &  1.2$\cdot10^{-3}$ & 1.2$\cdot10^{-4}$ & & 6.0$\cdot10^{-3}$ & 1.3$\cdot10^{-3}$ &  2.6$\cdot10^{-4}$\\
\hline\hline
\end{tabular}
\end{adjustbox}
\label{tab:inter_extrapolation}
\end{supptable*}